\documentclass[prl,twocolumn,groupedaddress,showpacs]{revtex4}

\usepackage{graphicx}
\usepackage{amsmath}
\usepackage{bm}

\newcommand{\be}{\begin{equation}}
\newcommand{\ee}{\end{equation}}
\newcommand{\bea}{\begin{eqnarray}}
\newcommand{\eea}{\end{eqnarray}}
\def\source#1#2#3#4{#1 {\bf #2}, #3 (#4)}
\def\book#1#2#3{{\it #1} (#2), #3}
\def\Sch{Schr\"{o}dinger}
\def\Eq#1{Eq. (\ref{#1})}
\def\half{\frac{1}{2}}

\def\eps{\varepsilon}
\def\epsc{\varepsilon_c}
\def\DY{\Delta Y}
\def\DYo{\Delta Y_o}
\def\DYc{\Delta Y_c}
\def\DE{\Delta E}
\def\DEo{\Delta E_o}
\def\DEc{\Delta E_c}
\def\Ymin{Y_{min}}
\def\lo{l_o}
\def\lt{l_t}
\def\FF{{\cal F}}
\def\QQ{{\cal Q}}
\def\kp2{\kappa^2}
\def\ga{\gamma}
\def\om#1{\omega_{#1}}
\def\omt#1{\tilde{\omega}_{#1}}
\def\ave#1{\langle #1 \rangle}

\begin{document}


\title{Quantum Effects in the Mechanical Properties of Suspended Nanomechanical Systems}


\author{S.M. Carr}
\author{W.E. Lawrence}
\author{M.N. Wybourne}
\affiliation{Department of Physics and Astronomy\\Dartmouth College\\Hanover, NH 03755\\}


\begin{abstract}
We explore the quantum aspects of an elastic bar 
supported at both ends and subject
to compression.  If strain rather than stress is held
fixed, the system remains stable beyond the 
buckling instability, supporting two potential minima.  
The classical equilibrium transverse displacement is
analogous to a Ginsburg-Landau order parameter, with 
strain playing the role of temperature.  We calculate
the quantum fluctuations about the classical value as 
a function of strain.  Excitation energies and quantum 
fluctuation amplitudes are compared for silicon beams
and carbon nanotubes.   
\end{abstract}

\maketitle

The continuing drive towards semiconductor device 
miniaturization and integration has resulted in 
fabrication and micromachining technologies that are 
capable of producing artificial structures with features 
approaching the ten nanometer length scale.  To go beyond 
this scale, naturally occurring and chemically organized 
structures are receiving much attention.  The availability 
of these top-down and bottom-up nanofabrication 
capabilities has initiated the new area of nanomechanics 
\cite{Roukes,Craighead,Park,Erbe,Cleland1} in which 
ultra small mechanical 
systems are used to explore both fundamental and 
applied phenomena.   Recently, two reports have 
appeared on two-state nanomechanical systems.  In one 
\cite{Rueckes}, crossed carbon nanotubes were suspended 
between supports and the suspended element was 
electrostatically flexed between two states.  In the 
second \cite{Cleland2}, it was proposed to use an 
electrostatically flexed cantilever to explore the 
possibility of tunneling in a nanomechanical system.

\begin{figure}[b]
	\includegraphics[width=8.5cm,height=7.5cm]{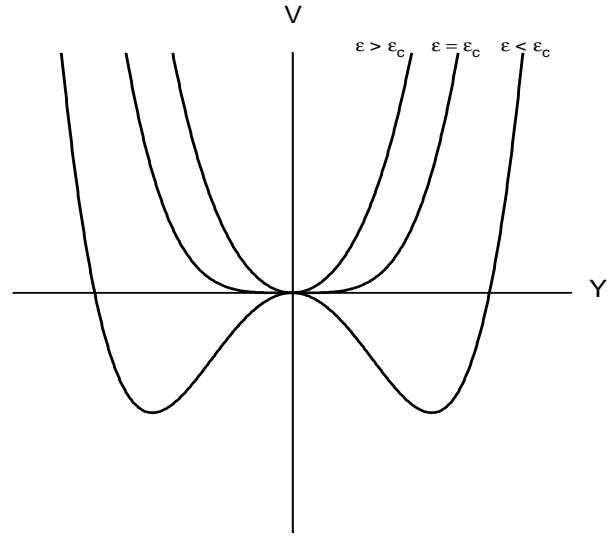}
	\caption{Potential energy $ V $ as a function of the fundamental mode displacement $ Y $.
	The shape of the potential energy is harmonic for $ \varepsilon > \varepsilon_{c} $, 
	quartic for $ \varepsilon = \varepsilon_{c} \equiv $ critical strain ($ \varepsilon_{c} < 0 $),
	and a double-well for $ \varepsilon < \varepsilon_{c} $.}
	\label{Fig1}
\end{figure}

In this Letter we discuss quantum effects in 
a two-state mechanical system 
that has a tunable, symmetric potential function.  This 
mechanical system has analogies to the superconducting 
interference device in which the first observation of a 
coherent superposition of macroscopically distinct 
states was recently reported \cite{Lukens}.
Specifically, we consider a suspended elastic bar under 
longitudinal compression.  The compression is used to 
adjust the potential energy for transverse displacements 
from the harmonic to the double-well regime, as shown in
Fig.~\ref{Fig1}, with 
strain playing a role analogous to temperature in a 
Ginzburg-Landau system.  
As the compressional strain is 
increased to the buckling instability \cite{Euler}, the 
frequency of the fundamental vibrational mode drops 
continuously to zero.  By controlling the separation 
between the ends of the bar, i.e. fixing the strain, the 
system remains stable beyond the instability and develops 
a double well potential for the transverse motion. Since 
both the well depth and asymmetry are tunable, a variety 
of quantum phenomena may be explored, including zero-point
fluctuations, tunneling, and coherent superpositions of
macroscopically distinct states. In the latter two cases, 
the system may provide a mechanical realization (at least 
in theory) of models studied in Refs. \cite{Caldeira} 
and \cite{Leggett}, respectively.  We have applied the 
model to suspended silicon beams and carbon nanotubes, 
and show that in both cases the quantum fluctuations 
in position approach 0.1 \AA; an order of magnitude
greater than the relaxed values.  We argue that 
compressing a carbon nanotube at low temperature may 
cause a crossover between the quantum and thermal 
fluctuation regimes.  Further, we suggest that tunneling
in this mechanical system may be difficult to achieve
because it will require precise regulation of the 
applied strain.

As a starting point, we consider the normal modes and
associated quantum properties of an elastic rectangular
bar of length $l$, width $w$~and thickness $d$ satisfying
($l>>w>d$), supported at both ends without strain.  We 
suppose that $d$~is smaller than $w$~so that transverse 
displacements $y(x,t)$ only occur in the ``$d$'' direction.  
The equation of motion for small displacements is \cite{LL}
\be
\label{DE1}
  \mu \ddot{y} + \FF \kp2 y^{(4)} = 0,
\ee
where $\mu = m/l$ is the mass per unit length and $\FF$~is 
the linear modulus (energy per unit length) of the bar.  
$\FF$ is related to 
the elastic modulus $\QQ$~of the material by $\FF = 
\QQ w d$.  The bending moment $\kappa$ is given by
$\kp2 = d^2/12$ for a bar of rectangular cross 
section.
 
The normal modes of the bar, 
$y(x,t) = y_n(x) \exp(\pm i \om{n} t)$, have the 
general spatial dependences 
\bea
\label{modefun}
 y_n(x)  = & a_n \cos k_n x + b_n \cosh k_n x, 
 \quad  & n = 1, 3, 5,\ldots     \\
 y_n(x)  = & a_n \sin k_n x + b_n \sinh k_n x,
  \quad   & n = 2, 4, 6,\ldots
\eea 
with $a_n/b_n$~and wavenumbers $k_n$~fixed by the
boundary conditions.  The boundary conditions for 
hinged endpoints,  $y(\pm l/2) = 0 = y''(\pm l/2)$, 
lead to normal modes with $b_n=0$,
$k_n=n\pi/l$, and (angular) frequencies
\be
\label{omn}
 \om{n} = \sqrt{{\QQ \over \rho}}~\kappa~k_n^2
  = \sqrt{{\QQ \over \rho}}~\kappa~
  ({n \pi \over l})^2, \quad
  (\hbox{hinged b.c.}).
\ee
Clamped endpoints have boundary conditions  
$y(\pm l/2) = 0 = y'(\pm l/2)$; both $a_n$ and $b_n$ 
are nonzero, and the normal mode frequencies are given 
to good approximation by replacing $n$ with ($n+\half$) 
in Eq. (\ref{omn}).  Because of their simplicity, we 
shall refer to hinged boundary conditions when specific 
results are quoted.

The mean square displacement of the bar (both 
zero-point and thermal) is an incoherent superposition 
of contributions from each normal mode.  At the center, 
only even-parity modes contribute, so that 
\be
\label{zeropt}
   \ave{(y(0))^2} = \sum_{odd~n} {\hbar \over 2m^* 
   \om{n}}~[1 + 2f(\hbar \om{n}/kT)],
\ee
where $f(x) = 1/(e^x-1)$ is the thermal excitation 
number of the $n$-th mode.  For hinged boundary  
conditions, $m^*=m/2$~exactly, whereas in the clamped 
case $m^*$~is slightly smaller and weakly mode 
dependent.  In either case the fundamental mode is 
responsible for more than half the total mean square
displacement.

Longitudinal compression of the bar moves the 
frequencies downward, with a corresponding increase 
in the zero-point motion.  Compressive or tensile
strain contributes the ``elastic'' potential energy 
$V_e = (\FF /2l_o)(\lt -l_o)^2$, where 
$\lt = \int dx \sqrt{1+(y')^2} \approx l + 
\half \int dx (y')^2$ is the total (dynamic) length 
of the bar, $l$ is the endpoint separation, and 
$l_o$ is the unstressed equilibrium length.
We subtract the static contribution 
$(\FF /2l_o)(l -l_o)^2$ since it contributes nothing
to the dynamics, and add the ``bending''contribution
$V_b \sim \int dx (y'')^2$ to get
\be
\label{VT}
 V[y(x)] = \half \int dx \left(\FF \kp2 (y'')^2 
 + \FF \eps (y')^2 \right) +
  {\FF \over 8 \lo} \Big(\int dx (y')^2 \Big)^2,
\ee
where $\eps \equiv (l-l_o)/l_o$ is the strain,
positive if tensile and negative if compressive.
From the Lagrangian, 
$L[y(x,t)] = (\mu/2) \int dx~(\dot{y})^2 - V[y(x)]$,
we find the equation of motion,
\be
\label{DE2} 
  \mu \ddot{y} + \FF \kp2 y^{(4)} -
   \FF \eps y'' - \half \FF
  \Big( \int dx' [y'(x')]^2 \Big) y'' = 0,
\ee 
which generalizes \Eq{DE1}.
The third term represents the tension induced by 
externally-imposed stretching, and the anharmonic 
fourth term is the enhancement of tension due to 
dynamic stretching; this term arises from the 
geometry of the system. 

In the harmonic regime where where the fourth term
can be neglected, the normal mode frequencies under
hinged boundary conditions are given by
\be
\label{omn2}
  \omt{n}^2 = \om{n}^2 \left(1 + \eps
  \Big({l \over n \pi \kappa}\Big)^2 \right),
\ee
where $\om{n}$ are the relaxed ($l=l_o$) frequencies 
of Eq. (\ref{omn}).  The zero-point fluctuations are 
given by Eq.~(\ref{zeropt}) with $\om{n}$~replaced by
$\omt{n}$.  Of course, the
harmonic approximation breaks down for the 
fundamental mode as we approach its critical strain,
\be
\label{crit}
  \epsc \equiv {l_c - l_o \over l_o} = 
   - \Big({\pi \kappa \over l}\Big)^2. 
\ee 
At critical strain, the effective potential for the
fundamental mode is purely quartic (see Fig.~\ref{Fig1}) whereas the higher 
modes remain harmonic in leading order, with the 
first harmonic frequency being reduced by about 13\% 
from its uncompressed value. 

To go beyond the harmonic approximation and calculate
the excitation energies and quantum fluctuations of the
fundamental mode, we consider the 
Hamiltonian
\be
\label{Hamiltonian1}
  H = {1 \over 2 \mu} \int dx~\Pi ^2 + V[y(x)],
\ee
where $\Pi(x,t) = \delta L/\delta \dot{y}(x,t) =\mu 
\dot{y}(x,t)$ is the canonical momentum.  In the
subcritical compression regime, a normal mode 
expansion of $H$ leads to a phonon description with
interactions arising from the anharmonic term.  These
interactions occur physically because phonons stretch 
the bar.  Even the
zero-point motion has a stretching effect, but this can 
be absorbed in the length parameter $l$.   Thus, at 
temperatures below the first harmonic threshhold, 
$kT < \hbar \omt{2}$, the anharmonic effect on the 
fundamental is its own self-interaction.  So the 
effective Hamiltonian for the fundamental mode, obtained
by taking the ground state expectation value in all
higher modes, is a quartic function of the 
``fundamental displacement'' $Y$, the Fourier 
component of the fundamental mode \cite{Fourier}. The 
quantum states and associated energy spectrum of 
fundamental vibrational states are then given by 
the \Sch~equation,
\be
  \left( -{\hbar^2 \over 2 m^*} {\partial^2 \over \partial Y^2} + {\alpha \over 2} Y^2 + 
  {\beta \over 4} Y^4 \right) \Psi(Y) = E \Psi(Y)
  \label{Sch1}
\ee
where $E_m$ and $\Psi_m(Y)$ are the energy eigenvalues
and eigenfunctions, and $-i \hbar \partial / \partial Y = P$ is the 
momentum operator canonically conjugate to $Y$.

\begin{figure}[b]
	\includegraphics[width=8.5cm,height=7.5cm]{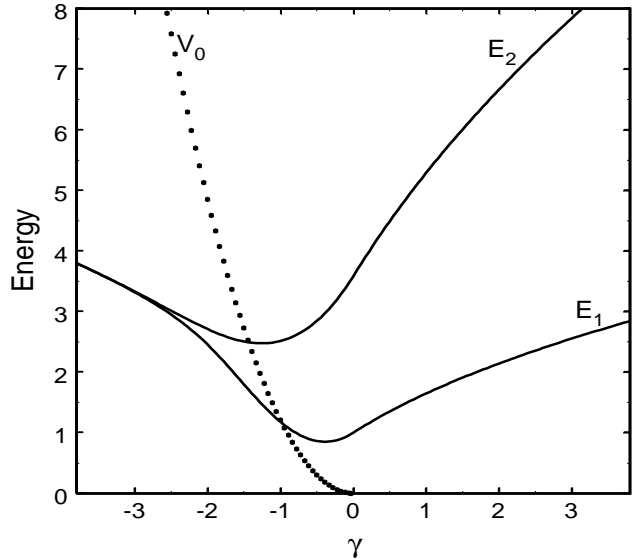}
	\caption{Ground and first excited state energies, $ E_{1} $ and $ E_2 $, of the fundamental
	vibrational mode as functions of the departure from critical strain $ \gamma $.  The dotted
	curve shows the barrier height dependence on $ \gamma $.}
	\label{Fig2}
\end{figure}

\begin{figure}[!]
	\includegraphics[width=8.5cm,height=7.5cm]{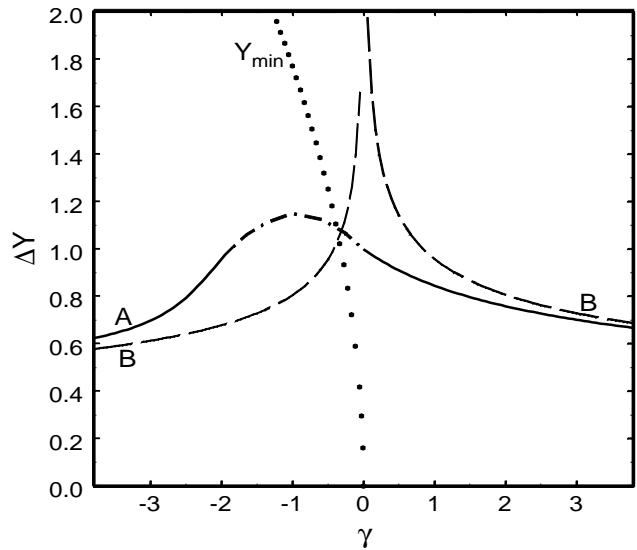}
	\caption{Curve A is the ground state fluctuation $\DY$, calculated 
	using the full quartic potential.  The solid 
	portions of the curve are obtained from the calculations; the 
	dot-dash region is a guide to the eye.  For comparison, curve B 
	shows $\DY$ obtained in the harmonic approximation.  The dotted 
	curve shows the position of the potential minimum, $\Ymin$.}
	\label{Fig3}
\end{figure}

The potential energy has the form of a Ginzburg-Landau
free energy \cite{GL}, with strain playing the role of 
temperature:
\be
\label{alpha}
   \alpha = m^* \omt{1}^2 = m^* \om{1}^2 
   \bigg({\epsc - \eps \over \epsc}\bigg).
\ee
The displacement $Y$ is analogous to the
order parameter, in that its classical equilibrium 
value vanishes below critical strain but takes a
nonzero value $Y \rightarrow \pm Y_{min}
= \pm \sqrt{|\alpha|/\beta}$ above it, 
breaking the reflection symmetry of the Hamiltonian.  
Of course the quantum mechanical ground state has 
$\ave{Y} = 0$, but sufficiently far into the 
double-well regime, this ground state is a 
superposition of macroscopically distinct states.
Thus, monitoring the position of the bar on a time
scale less than the tunneling time would yield results 
clustered about just one of the potential minima, 
$Y_{min}$~or $-Y_{min}$, not both.  With this in 
mind, we plot  the ground and first excited state 
energies in Fig.~\ref{Fig2}, and in Fig.~\ref{Fig3} the ground 
state quantum fluctuations $\DY$, as functions 
of the strain near its critical value.  These 
are plotted in dimensionless energy and length 
units, $E/E_c$ and $\DY/\DYc$, where 
$E_c$ and $\DYc$ are the ground state values at 
critical strain,
\be
\label{EcYc1}
 E_c = 0.42\bigg({\hbar^2 \over m^*}\bigg)^{2/3}
 \beta^{1/3}  \quad \hbox{and} 
 \quad \DYc = 0.68\bigg({\hbar^2 
  \over m^* \beta}\bigg)^{1/6},
\ee
and $\ga$~is the departure from critical strain, 
\be
\label{gamma}
  \ga = {\epsc - \eps \over \epsc} 
    \bigg({1.6 \kappa \over \DY_c}\bigg)^2,
\ee
scaled so that $\ga = -1$~when the barrier height 
$V_o = \alpha^2/4 \beta$ is equal to $E_1$.
\begin{table*}
\caption{Table of excitation energies $\DE =
E_2 - E_1$ and $rms$ midpoint fluctuations $\Delta Y$ for Si bars 
(linear dimensions $l$, $d$, $w$), and C nanotubes of length $l$ and 
outer (inner) diameters $d_2$ ($d_1$). The Young's modulus and density 
of Si are $Q = 130$ GPa and $\rho = 5000$ kg/m$^3$.  The values taken for 
C nanotubes are \cite{Treacy} $Q = 1.8$ TPa and  $\rho = 2150$ kg/m$^3$.
Energies $\DE$~are given in temperature units (note that 1GHz = 48mK).}
\label{Table}
$$
\vbox{
  \halign{
  # \hfil \quad & \hfil # \hfil \quad & \hfil # \hfil \quad & \hfil # \hfil
   \quad & \hfil # \hfil \quad & \hfil # \hfil \quad & \hfil # \hfil     \cr                                                         \cr
\noalign{\hrule}
\noalign{\smallskip}
\noalign{\hrule}
\noalign{\medskip}
   & Si bar & Si bar &  & C nanotube & C nanotube   \cr
\noalign{\medskip}
\noalign{\hrule}
\noalign{\smallskip}
 $l$(nm) & 500 & 50 & $l$(nm) & 500 & 50                       \cr
 $d$(nm)  & 10 & 5 & $d_2$(nm) & 10 & 5                            \cr
 $w$(nm)  & 20 & 10 & $d_1$(nm) & 5 & 1                            \cr
\noalign{\smallskip}
\noalign{\hrule}
\noalign{\smallskip} 
\noalign{\hrule}
\noalign{\smallskip} 
  & & & relaxed & &                                             \cr
 $\DEo$(mK) & 4.4 & 222 &  & 24.4 & 1100                       \cr
 $\DYo$(\AA) & .0061 & .0054 &  & .0093 & .0082                \cr
\noalign{\smallskip}
\noalign{\hrule}
\noalign{\smallskip} 
  & & & critical & &                                            \cr
 $\DEc$(mK) & .027 & 2.0 &  & 0.21 & 14.5                      \cr
 $\DYc$(\AA) & .055 & .041 &  & .073 & .051                    \cr            
\noalign{\medskip}
\noalign{\hrule}
\noalign{\smallskip}
\noalign{\hrule}
}}
$$
\end{table*}
In defining $\DY$, care should be taken to account 
for the onset of a nonzero mean displacement as one 
moves into the double well regime.  In the
subcritical and near-critical regimes, the usual 
definition $\DY^2 = \ave{Y^2} - \ave{Y}^2$ surely
applies, where brackets indicate ground state 
expectation values and $\ave{Y} = 0$.  However, 
the comments above suggest that deeper in the 
double well regime, a more sensible practical 
definition is $\DY^2 = \ave{min(Y\pm\Ymin)^2}$, 
the $rms$~departure from the nearest potential 
minimum. Fig.~\ref{Fig3} shows that quantum fluctuations
defined in this way become much less than the
well separation at fairly modest (negative) values 
of $\ga$.  In the region $\ga \sim -1$, the latter 
definition of $\DY^2$ loses its physical meaning
and larger fluctuations may be expected, as 
suggested by the dot-dash line on Fig.~\ref{Fig3} which 
is only a guide to the eye. 

A useful perspective is provided by comparing the 
exact quantum fluctuations with those given by the 
harmonic approximation.  Indeed, making this 
approximation in the double well regime requires 
that one of the wells is ignored while fluctuations 
are defined about the minimum of the other, 
which clearly will become invalid as $\DY$
approaches $\Ymin$ in the near-critical regime.
As shown in Fig.~\ref{Fig3}, this approximation 
predicts that the fluctuation $\DY$ diverges as 
$|\eps - \epsc|^{-1/4}$.  
It is nonetheless a very good approximation outside a 
small region around the 
critical point (roughly $-6 < \gamma < 3$), where the 
divergence is prevented by the quartic term in the 
potential energy. 

To address the magnitude of quantum fluctuations in 
real systems, Table~\ref{Table} lists the first excitation energies 
$\DE = E_2 - E_1$~and ground state (quantum) fluctuations 
$\DY$ \cite{Fourier} for rectangular silicon bars and 
cylindrical multiwalled carbon nanotubes. 
Numbers are given for two cases - the critical and 
the relaxed (uncompressed) states.  The remaining 
\Sch~equation parameters are calculated using hinged 
boundary conditions, with the results 
\be
\label{pms}
 m^* = m/2 \hskip1truecm \hbox{and}
 \hskip1truecm \beta = m^* (\om{1}/2 \kappa)^2.
\ee
The larger tubes have dimensions typical of the 
multiwalled tubes whose vibrational properties were 
studied by Treacy {\it et al.} \cite{Treacy}.  The 
smaller tube is probably the smallest that would support 
buckling and retain its elastic integrity \cite{Lourie}.  
Nanotube frequencies were found using \Eq{omn} with 
$\kp2 = (d_2^2 + d_1^2)/16$, where $d_1$ and $d_2$ are
inner and outer diameters.  In the critical and double
well regimes, because of cylindrical symmetry,
the nanotube states will be described by a Mexican hat 
potential rather than a double well. In this case, 
\Eq{Sch1} still applies, with $Y$ replaced by a 
two-component vector. In order to make a comparison to 
silicon bars, the nanotube entries in Table~\ref{Table} refer 
to a single Cartesian component of $Y$.  Fig.~\ref{Fig3} 
refers specifically to one-dimensional transverse 
motion; the Mexican hat potential would have its main 
effect in the negative $\ga$ regime where the first 
excitation energy $E_2 - E_1$ would fall less 
rapidly because the excitations are rotational 
rather than tunneling in character.  

For all systems shown, the zero-point fluctuations 
predicted in the relaxed states are enhanced by nearly 
an order of magnitude by applying critical compression.
On the other hand, the excitation energies are all 
reduced by about two orders of magnitude, so that thermal
fluctuations can swamp the quantum contributions.   
On the other hand, 
the large differences between the relaxed and compressed 
energy scales suggests that the relaxed systems could be 
supercooled by compression toward their 
critical points.  For example, the smaller carbon nanotube 
could be prepared initially very close to its ground state 
by cooling to 500 mK.  Critical compression without heat 
transfer would then cool the tube to a few mK, at the
same time enhancing its zero-point motion by a factor of
about 6 (see Table~\ref{Table}).  Subsequent equilibration to 500 mK 
would then bring the tube to a ``classical'' equilibrium 
state with thermal fluctuation $\DY_t \approx 0.25$~\AA\,
\cite{thermal}, a further factor of 5 enhancement. 

Finally, we comment on the possibility of observing 
tunneling in nanomechanical systems such as those 
considered here. 
We can roughly estimate the number of bound states with 
energy below the top of the barrier; it is of the order
of the barrier height $V_o$ divided by the level 
spacing $\hbar \omt{1}$. From Eqs. \ref{omn}, \ref{alpha} 
and \ref{pms} we find
\be
\label{bound}
  N \sim \bigg({\kappa \over \DY_o}\bigg)^2
  \bigg({\epsc - \eps \over \epsc}\bigg)^{3/2}.
\ee
Taking the smaller bar, if we go to twice the critical 
compression, $\eps = 2\epsc$, then $N \sim 3 \times 10^6$. 
In order to tune the potential to hold about 10 bound 
states in each well, one would have to apply strain with 
extreme delicacy, $\eps - \epsc \sim 10^{-4} \epsc$.  
Controlling the strain to this precision for sufficient 
time to identify tunneling, as distinct from thermal or
other noise, will be difficult.  Thus, while the 
observation of tunneling will likely be very challenging, 
the prospect of exploring tunable quantum fluctuations in 
this system, and the connection to Ginzburg-Landau theory, 
are intriguing.


\begin{thebibliography}{99}
\bibitem{Fourier} We scale $Y$ to be equal to
   the central displacement in the fundamental mode, so
   that $y(x) = Y y_1(x)/y_1(0)$. The {\it rms} fluctuations are
   then $(\DY)^2 = \ave{(y(0))^2}$.
\bibitem{GL} This is not a close analogy:  The potential 
  applies to the single degree of freedom $Y$, not to 
  the field $y(x)$, so that the usual ``stiffness''
  term $\sim (\partial / \partial x)^{2}$ of Ginzburg-Landau theory is 
  absent.  The present $(\partial / \partial Y)^{2}$~term is a quantum 
  mechanical operator usually absent in Ginzburg-Landau theory.
\bibitem{thermal} Calculated using the generalized 
   equipartition result that degrees of freedom appearing
   quartically in the Hamiltonian contribute $kT/4$
   to the internal energy.
\end{thebibliography}
\end{document}